\documentclass[12pt]{article} 
\newcommand{\be}{\begin{equation}}
\newcommand{\ee}{\end{equation}}
\newcommand{\ba}{\begin{array}}
\newcommand{\ea}{\end{array}}

\newcommand{\bea}{\begin{eqnarray*}}
\newcommand{\eea}{\end{eqnarray*}}
\newcommand{\bean}{\begin{eqnarray}}
\newcommand{\eean}{\end{eqnarray}}
\newcommand{\proof}{\vspace{1ex}\noindent{\em Proof}. \ }
\def\ds{\displaystyle}

\newtheorem{lemma}{Lemma}[section]

\newtheorem{definition}{Definition}[section]
\newtheorem{theorem}{Theorem}[section]
\newtheorem{corollary}{Corollary}[section]
\newtheorem{proposition}{Proposition}[section]
\usepackage{amsmath, amssymb}
\usepackage{latexsym}
\newcommand{\R}{\mathbb{R}}

\def\Box{\leavevmode\vbox{\hrule
     \hbox{\vrule\kern5pt\vbox{\kern5pt}%
           \vrule}\hrule}}
\renewcommand{\square}{\hfill$\Box$}
\begin{document}
\title{Inverse Conduction Problem for a Parabolic Equation using a Boundary Integral Method}

\author{Christian Daveau \thanks{
D\'epartement de Math\'ematiques, Site Saint-Martin II, BP 222, \&
Universit\'e de Cergy-Pontoise, 95302 Cergy-Pontoise Cedex, France
(Email: christian.daveau@math.u-cergy.fr).} \and Abdessatar
Khelifi
\thanks{ D\'epartement de Math\'ematiques \& Informatique
Facult\'e des Sciences, 7021 Zarzouna - Bizerte, Tunisia
(Email:abdessatar.khelifi@fsb.rnu.tn).} \and M. Nour. Shamma
\thanks{Department of mathematics \& T. College,
Al Gassim University, P. O. Box 53, Al-Rass Province, Kingdom of
Saudi Arabia(Email:shamman01@yahoo.com).}}

\maketitle \abstract{ In this paper, a boundary integral method is
used to solve an inverse linear heat conduction problem in
two-dimensional bounded domain. An inverse problem of measuring
the heat flux from partial (on part of the boundary) dynamic
boundary measurements is considered. An algorithm is given by
using the fundamental solution. }\\



\section{Introduction}
This paper is devoted to an inverse problem for a type of
parabolic PDEs in a bounded two-dimensional domain. Here we
consider initial boundary value problems for the heat equations by
using a boundary integral approaches  The inverse heat conduction
problem arising in most thermal manufacturing processes has
recently attracted much attention
\cite{ang,cheng,cheng-F,jia,ramm1,ramm2,wang}. The typical case is
the determination of the heat flux on an inaccessible boundary
through measurements on an accessible boundary or inside the
domain. Similar problems to ours have been studied by many
authors, for example in one dimensional space we can refer to an
approximate inverse method \cite{jia,jonas}, a boundary element
method \cite{lesnic}, a fundamental solution method
\cite{bogomly,evans98} and some other method \cite{y-shen}. In
this paper, we use the boundary integral method to solve this
problem. This method uses the prescribed initial and boundary
data, together with the fundamental solution of a given
differential equation defined in some bounded domain, and we
construct integral equations on the boundary of ­. In our case,
the solution to the integral equation is a single layer potential.
By the boundary integral equation one can obtain the unknown
kernel, and the solution to the given problem will be obtained by
integrating the product of the fundamental solution and the
unknown kernel over the boundary. The advantage of our approach is
that the computation can be limited to the boundary, which reduces
the problem from two dimensions to one dimension. As a result of
the reduction, we may expect substantial savings in computer time
and memory. The work outlined below is based on the use of single
layer potentials. Ammari and Kang used boundary integral method to
solve inverse conductivity problem and related problems
\cite{A-K}. In \cite{A-K}, both the single layer potential and the
double layer potential are used. In this article, our boundary
integral method is based on the result of \cite{saranen1}, which
gives a representation formula for the heat conduction problem
with Neumann boundary condition. The equation is assumed to be
homogeneous. The outline of our paper is as follows:...

\section{Problem formulation}

Let $\Omega\subset\R^2$ be a bounded domain of $\R^2$. We denote
$\partial\Omega$ the boundary set assumed to be of class $C^1$. We
denote by $\nu$ the outward unit normal to $\Omega$ on
$\partial\Omega$. Let $T>0$. We consider the following homogeneous
heat equation: \be\label{bvp}
\begin{array}{cc}
  \partial_tu-\Delta u=0, & (x,t)\in \Omega\times[0,T] \\
  u(x,0)=u_0(x), & x\in\Omega \\
   u(x,t)=g(x,t), & (x,t)\in \partial\Omega\times[0,T]\\
  \frac{\partial u}{\partial\nu}(x,t)=\phi(x,t), & (x,t)\in\partial\Omega\times[0,T] \\
\end{array}
\ee

where $u(x, t)$ is the temperature function, $u_0(x)$ the initial
data, $g(x,t)$ is a suitably prescribed function, $\phi(x,t)$ is
the unknown {\it heat flux} and $\ds\partial_tu=\frac{\partial
u}{\partial t}$ is the rate of change of temperature at a point
over time. Notice that $\ds \partial_\nu u=\frac{\partial u}{\partial\nu}=\nabla u\cdot \nu$.\\

Let $\gamma(\zeta):[0,1]\to \R^{2}$ be an analytic, $1$-periodic
 function.
We assume that the tangential derivative has the positive length
$|\gamma^{\prime}(\zeta)|>0$ for all $0\leq \zeta\leq 1$.\\

Throuout this paper we suppose that the closed smooth curve
$\partial\Omega$ in $\R^2$ is parameterized by the function
$\gamma(\zeta)$ as follows :
\begin{equation}\label{curve1}\partial\Omega=\{x=\gamma(\zeta),\quad
\zeta\in[0,1] \}.\end{equation}

The present paper proposes a boundary integral method for the
numerical solution of the two-dimensional problem defined by
(\ref{bvp}). Our problem can be stated as follows:

\subsection*{The inverse problem}
Let $\Gamma\subset\partial\Omega$ denote a measurable smooth
connected part of the boundary $\partial\Omega$
($\Gamma=\partial\Omega$ or not). The aim of this paper is to
determine the heat flux $\ds \phi(x,t)=\frac{\partial
u}{\partial\nu}|_{\Gamma}(x,t)$ from measurements of :
\[
u(x,t)\quad \mbox{ on } \Gamma\times[0,T].
\]
For this purpose, we develop a boundary integral method .... as
will be described in the next section.

\section{Boundary integral method}
In this section we consider the basic boundary integral approaches
for the solution of the initial boundary value problem
(\ref{bvp}). As in the time-independent case there are two main
types of approaches, namely the direct method based on the
representation coming from Green's formula and the indirect or
layer methods (see for example \cite{saranen}). We begin with the
fundamental solution of the heat equation. In several spatial
variables, the Green's function is a solution of the initial value
problem (see for example \cite{evans98}): \be\label{green1}
  \partial_tG-\Delta G=0,\quad \mbox{and } G(x,t=0)=\delta(x)
\ee where $\delta$ is the Dirac delta function. The solution to
this problem in $\R^n$ ($n\geq 1$) is the fundamental solution :
\[
\ds G(x,t)=\frac{H(t)}{(4\pi t)^{n/2}}\exp(-\frac{|x|^2}{4t}),
\]
where $H(t)$ is the Heaviside function and $|x|=(x_1^2+x_2^2+\cdots+x_n^2)^{1/2}$ for $x=(x_1,x_2,\cdots,x_n)\in\R^n$.\\
Let $\mathcal{S}$ and $\mathcal{D}$ denote the classical
single-layer and double-layer heat potentials : \[
(\mathcal{S}q)(x,t)=\int_{0}^{t}\int_{\partial\Omega}G(x-y,t-s)q(y,s)~d\sigma(y)ds,\]
\[
(\mathcal{D}\varphi)(x,t)=\int_{0}^{t}\int_{\partial\Omega}\partial_{\nu(y)}G(x-y,t-s)\varphi(y,s)~d\sigma(y)ds,\]
and similarly we define $\mathcal{D}^{\prime}$ as the spatial
adjoint of the double-layer $\mathcal{D}$ and $\mathcal{H}$ the
hyper singular heat operators as follows:
\[
(\mathcal{D}^{\prime} q
)(x,t)=\int_{0}^{t}\int_{\partial\Omega}\partial_{\nu(x)}G(x-y,t-s)q(y,s)~d\sigma(y)ds,\quad
\mbox{for } (x,t)\in\partial\Omega\times[0,T],\] and
\[
(\mathcal{H}\varphi)(x,t)=-\int_{0}^{t}\int_{\partial\Omega}\frac{\partial^2
G(x-y,t-s)
}{\partial_{\nu(x)}\partial_{\nu(y)}}\varphi(y,s)~d\sigma(y)ds,\quad
\mbox{for } (x,t)\in\partial\Omega\times[0,T];\] where $q$ and
$\varphi$ sufficiently smooth functions.\\
In terms of the operators $\mathcal{S}$ and $\mathcal{D}$, we
introduce the following heat potential
\begin{equation}\label{poten1}
u=\mathcal{S}q - \mathcal{D}\varphi\quad  \mbox{in
}\Omega\times[0,T],
\end{equation}
where $u$ is the solution of (\ref{bvp}). Now, by using the well
known boundary behavior properties of single-layer and the
double-layer heat potential the following result holds.
\begin{proposition}\label{prop-integ1}
Let $u$ be the solution of the problem (\ref{bvp}). Then, in both
cases homogeneous or inhomogeneous initial data $u_{0}$, the
Cauchy data
$(u|_{\partial\Omega\times[0,T]},\partial_{\nu}u|_{\partial\Omega\times[0,T]})$
satisfy : \begin{equation}\label{poten2} u = \mathcal{S}q +
(\frac{1}{2}I-\mathcal{D})\varphi\quad \mbox{ in }
\partial\Omega\times[0,T]\end{equation}
\begin{equation}\label{poten3} \partial_{\nu}u = (\frac{1}{2}I+\mathcal{D}^{\prime})q+\mathcal{H}\varphi \quad \mbox{ in }
   \partial\Omega\times[0,T],
\end{equation}
where the functions $q$ and $\varphi$ are given as in
(\ref{poten1}).
\end{proposition}
\proof For the case of inhomogeneous initial data $u_{0}\equiv 0$
one can get the above results by using the well known boundary
behavior properties of single-layer and the double-layer heat
potential together with their normal derivatives. But, for
inhomogeneous initial data $u_{0}\neq 0$, the situation needs more
explanations. It can be seen that a function
\[
\ds v(x,t)=\int_{0}^{t}\int_{\partial\Omega}G(x-y,t)u_{0}(y)~dy
\]
satisfies
\[
v_t-\Delta v=0\] and
\[
\lim_{t\to 0^+}v(x,t)=u_0(x).\] Setting $w=u-v$, we get
\begin{equation}
\begin{array}{cc}
  \partial_tw-\Delta w=0, & (x,t)\in \Omega\times[0,T] \\
  w(x,0)=0, & x\in\Omega \\
 w(x,t)=g(x,t)
  -v(x,t), & (x,t)\in\partial\Omega\times[0,T], \\
  \frac{\partial w}{\partial\nu}(x,t)=\phi(x,t)
  -\frac{\partial v}{\partial\nu}|(x,t), & (x,t)\in\partial\Omega\times[0,T] .\\
\end{array}
\end{equation}
Since $w$ is the solution of (\ref{bvp}) with homogeneous initial
data, it can be solved to find the integral relations
(\ref{poten2})-(\ref{poten3}). Thus we can solve the problem and
get our results by the superpositions of $v$ and the solution of
homogenous initial data. \square

Now, we introduce the following notations and the anisotropic
Sobolev space to be used in the squeal. A rather comprehensive of
the basic presentation of these Sobolev spaces the reader can
see~\cite{lions,saranen}. For given $r,p\geq 0$, we have the space
\[
H^{r,p}(\Omega\times[0,T])=L^2([0,T];H^r(\Omega))\cap
H^p([0,T];L^2(\Omega)).
\]
The space $H^{r,p}(\partial\Omega\times[0,T])$ is defined
analogously by replacing $\Omega$ by $\partial\Omega$. Moreover
the following subspace of $H^{r,p}(\partial\Omega\times[0,T])$ is
well defined
\[
\hbar^{r,p}(\partial\Omega\times[0,T])=\{
v=w|_{\partial\Omega\times[0,T]}: w\in
H^{r,p}(\Omega\times\R),\quad w(\cdot,t)=0,t<0 \}.
\]
The norm of the space $H^{r,p}(\Omega\times[0,T])$ is denoted by
$\|\cdot\|_{r,p;\Omega\times[0,T]}$ and the norm of
$\hbar^{r,p}(\partial\Omega\times[0,T])$ is denoted by
$\|\cdot\|_{r,p;\partial\Omega\times[0,T]}$.\\
Next, introducing the useful spaces $$\mathcal{E}=\{v\in
H^{2,1}(\Omega\times[0,T]):\quad (\Delta+\partial_{t})v\in
L^{2}(\Omega\times[0,T]),\partial_{\nu}v|_{\partial\Omega\times[0,T]}\equiv
0,v(\cdot,T)|_{\Omega}\equiv 0 \},$$
$$
\mathcal{X}=\hbar^{1/2,1/4}(\partial\Omega\times[0,T])\quad
\mbox{and the associate dual space
 }\mathcal{X}^{\prime}=\hbar^{-1/2,-1/4}(\partial\Omega\times[0,T]).
$$ Then the following
definition appears.
\begin{definition}\label{def1}
For given $f\in \mathcal{X}^{\prime}$, we say that $u$ is a {\it
weak solution} of (\ref{bvp}) if $u\in
H^{1,1/2}(\Omega\times[0,T])$ and satisfies the following duality
product:
\[
\langle u,\Delta \psi+\partial_{t}\psi \rangle=-\langle f,\psi
\rangle_{\partial\Omega\times[0,T]},\] for $\psi\in\mathcal{E}$.
\end{definition}
In terms of the last notations, the following mapping properties
of the single- and double-layer heat operators holds.
\begin{lemma}\label{lem-mmapping}
Let the operators $\mathcal{S}$ and $\mathcal{D}$ be defined as in
section 2. Then the followings hold:
\begin{itemize}
  \item [(1)] The single-layer heat operator $\mathcal{S}:\hbar^{r,1/2r}(\partial\Omega\times[0,T])\to
  \hbar^{r+1,1/2(r+1)}(\partial\Omega\times[0,T])$ is an
  isomorphism for all $r\geq -1/2$.
  \item [(2)] The operator $\frac{1}{2}I+D:\mathcal{X}^{\prime}\to
  \mathcal{X}^{\prime}$ is an isomorphism.
\end{itemize}
\end{lemma}
\proof The claim (1) can be inspired directly from Theorem 4.3 in
\cite{saranen}. The claim (2) follows by a little modification
from \cite{costabel1}. \square

Now we're ready to prove the following result.
 \begin{theorem}\label{thm1}
 Assume that $\phi\in \mathcal{X}^{\prime}$, Then the function $u\in H^{1,1/2}(\Omega\times[0,T])$ is a weak solution of (\ref{bvp}) if and
 only if $u$ has the representation (\ref{poten1}) such that
 $\varphi\in \mathcal{X}^{\prime}$ solves the equations (\ref{poten2})-(\ref{poten3}).
 \end{theorem}
 \proof
 Let $\phi\in \mathcal{X}^{\prime}$, and let $\varphi\in \mathcal{X}^{\prime}$ be the
 unique solution of (\ref{poten2})-(\ref{poten3}). Then, by combining relation (\ref{poten2}) and the direct representation (\ref{poten1}),
  we may get the following boundary integral equation of the second kind :
  \begin{equation}\label{biesk1}
  (\frac{1}{2}I+\mathcal{D})\varphi=\mathcal{S}\phi.
  \end{equation}
 Since the set $\mathbf{D}(\partial\Omega\times[0,T])$ is dense
 in the space $\mathcal{X}^{\prime}$, we can choose a sequence $\varphi_n\in
 \mathbf{D}(\partial\Omega\times[0,T])$ such that
 $\varphi_n\to\varphi$ in $\mathcal{X}^{\prime}$, and so the following sequence
 $\phi_n=\mathcal{S}^{-1}(\frac{1}{2}I+\mathcal{D})\varphi_n$ is well defined. By the mapping properties
 of $\mathcal{S}$ and of the operator
 $\frac{1}{2}I+\mathcal{D}$ found in Lemma \ref{lem-mmapping}, the function $\phi_n$ is
 also a smooth function of $\partial\Omega\times[0,T]$; moreover
 we have $\phi_n\to \phi$ in $\mathcal{X}^{\prime}$.\\ Now, let $u_n$ be the
 corresponding classical potential
 \[
 u_n=\mathcal{S}\varphi_n-\mathcal{D}\phi_n,
 \]
which by the construction satisfies
\[
\partial_{\nu}u_n|_{\partial\Omega\times[0,T]}=\phi_n.
\]
Obviously, $u_n$ is a weak solution of (\ref{bvp}) with the
Neumann data $\phi_n$. On the other hand, As done for the
Dirichlet-type initial boundary value problem in \cite{saranen},
we can conclude for our problem that the mapping $\phi\to u$, is
continuous and we have $u\in H^{1,1/2}(\Omega\times[0,T])$ such
that
\begin{equation}\label{cont1}
\|u\|_{1,1/2,\Omega\times[0,T]}\leq c
\|\phi\|_{1/2,1/4,\partial\Omega\times[0,T]},\quad \mbox{where } c
\quad \mbox{is a positive constant}.
\end{equation}
Hence, by the continuity (\ref{cont1}) we have the convergence
$u_n\to u$ in $H^{1,1/2}(\Omega\times[0,T])$. Next, we define
$\psi=\mathcal{S}\varphi-\mathcal{D}\phi$ in $\Omega\times[0,T]$,
then for all $v\in \mathbf{D}(\Omega\times[0,T])$ it follows that
\[
\langle u,v\rangle=\lim_{n\to\infty}\langle u_n,v
\rangle=\lim_{n\to\infty}\big(\langle
\mathcal{S}\varphi_n,v\rangle-\langle
\mathcal{D}\phi_n,v\rangle\big) =\langle \psi,v\rangle,
\]
which implies that $u=\psi=\mathcal{S}\varphi-\mathcal{D}\phi.$
\square

\section{Numerical scheme for the inverse problem}
In this section we propose a numerical method to solve our inverse
problem. The numerical method is based on the boundary integral
equation in Proposition \ref{prop1}. As the measured data for
inverse problem, the numerical data obtained by solving the direct
problem can be used. The inverse problem is then to solve the
following problem:
 \be\label{bvp-inv1}
\begin{array}{cc}
  \partial_tu-\Delta u=0, & (x,t)\in \Omega\times[0,T] \\
  u(x,0)=0, & x\in\Omega \\
  u(x,t)=g(x,t), & (x,t)\in\partial\Omega\times[0,T] \\
  \frac{\partial u}{\partial\nu}(x,t)=\phi(x,t), & (x,t)\in\partial\Omega\times[0,T]. \\
\end{array}
\ee We use the following equation given by (\ref{poten1})
\begin{equation}\label{poten1-inv1}
u|_{\partial\Omega\times[0,T]}=\mathcal{S}q -
\mathcal{D}\varphi\quad \mbox{in }\Omega\times[0,T],
\end{equation}
together with the relations (\ref{poten2})-(\ref{poten3}) to solve
the above problem.\\
To give our numerical numerical method, we introduce the following
result.
\begin{lemma}\label{lem-integ-num1}
Let $\varphi=[u]_{\partial\Omega\times[0,T]}$, $u$ solution of
(\ref{bvp-inv1}). In term of the hyper-singular heat operator the
heat flux is solution of :
\begin{equation}\label{integ-num1}
\mathcal{H}\varphi=\phi,
\end{equation}
where $[u]_{\partial\Omega\times[0,T]}$ means the jump of the
function $u$ via the boundary $\partial\Omega\times[0,T]$.
\end{lemma}
The boundary integral equation of the first kind given in Lemma
\ref{lem-integ-num1} is deduced from a normal derivative applied
to the double layer representation which itself given by managing
relation (\ref{poten1}) into relations given in Proposition
\ref{prop-integ1}.\\

To proceed with our numerical scheme, we may follows two cases,
measure on the boundary $\partial\Omega$ and measure on a smooth
connected subset of the boundary.

\subsection{Measure on the boundary $\partial\Omega$}
The aim of this section is to reconstruct $\phi(x,t)$ from
measurements of $u(x,t)$ on the boundary $\partial\Omega\times(0,
T)$. For this purpose, we develop the subdivision of $[0,1]$ :
\begin{equation}\label{subdiv1}
\zeta_0=0,\zeta_i=\zeta_0+ih,i= 1,2,\cdots,N,\quad \mbox{where }
N\quad \mbox{is an integer and }h=1/N.
\end{equation}
Analogously, we assume the subdivision of $[0,T]$ :
\begin{equation}\label{subdiv2}
t_0=0,t_j=t_0+jh^{\prime},j= 1,2,\cdots,N^{\prime} \quad
\mbox{where }N^{\prime}\quad \mbox{is an integer and
}h^{\prime}=T/N^{\prime}.
\end{equation}

Then, by using Lemma \ref{lem-integ-num1} the following main
result follows.
\begin{theorem}\label{thm-schem1}
Let $g\in \mathcal{X}$ be a given function and $u\in
H^{1,1/2}(\Omega\times[0,T])$ be the solution of (\ref{bvp-inv1}).
Assume that we have the subdivisions
(\ref{subdiv1})-(\ref{subdiv2}). Suppose that the heat flux of the
problem (\ref{bvp-inv1}) is continuous up to the inner side of the
inaccessible boundary $\partial\Omega$, then the unknown data
$\phi$ can be recovered by the following discrete scheme
\begin{equation}\label{schem1}
\ds\phi(\gamma(\zeta_i),t_j)=\frac{hh^{\prime}}{4}\sum_{k=1}^{N}\sum_{l=1}^{N^{\prime}}\frac{g(\gamma(\zeta_k),t_l)}{(t_j-
t_l)^2}\frac{\gamma^{\prime}(\zeta_i)}{|\gamma^{\prime}(\zeta_i)|}\cdot\big[-\gamma^{\prime}(\zeta_k)+
\end{equation}
\[
2\big(\gamma^{\prime}(\zeta_k)\cdot(\gamma(\zeta_k)
-\gamma(\zeta_i))\big)\frac{(\gamma(\zeta_i)-\gamma(\zeta_k))}{t_j-t_l}\big]\exp(-\frac{|\gamma(\zeta_i)-\gamma(\zeta_k)|^2}{(t_j-t_l)}),
\]
for  $i= 1,2,\cdots,N;\quad j= 1,2,\cdots,N^{\prime}$ and $\gamma$
is given by Section 2.
\end{theorem}
\proof Let $u$ be the solution of the inverse heat problem
(\ref{bvp-inv1}) and inserting the expression of the
hyper-singular operator $\mathcal{H}$ into relation
(\ref{integ-num1}), we get
$$
\phi(x,t)=-\int_{0}^{t}\int_{\partial\Omega}\frac{\partial^2
G(x-y,t-s)
}{\partial\nu(x)\partial\nu(y)}g(y,s)~d\sigma(y)ds,\quad \mbox{for
} (x,t)\in\partial\Omega\times[0,T].
$$
By change of variable $x=\gamma(\zeta), \zeta\in[0,1]$, we write
\begin{equation}\label{integ-num2}
\phi(\gamma(\zeta),t)=-\int_{0}^{t}\int_{0}^{1}\frac{\partial^2 G
}{\partial\nu(x)\partial\nu(y)}(\gamma(\zeta)-\gamma(\zeta^{\prime}),t-s)g(y,s)|\gamma^{\prime}(\zeta^{\prime})|~d\zeta^{\prime}ds,\quad
\mbox{for } (\zeta,t)\in[0,1]\times[0,T].
\end{equation}
Then according to (\ref{subdiv1})-(\ref{subdiv2}), we can
discretize relation (\ref{integ-num2}) as follows :
\begin{equation}\label{integ-num3}
\phi(\gamma(\zeta_i),t_j)=-hh^{\prime}\sum_{k=1}^{N}\sum_{l=1}^{N^{\prime}}\frac{\partial^2G}{\partial\nu(x)\partial\nu(y)}(\gamma(\zeta_i)-
\gamma(\zeta_k),t_j-t_l)g(\gamma(\zeta_k),t_l)|\gamma^{\prime}(\zeta_k)|.
\end{equation}
On the other hand the normal derivative of the fundamental
solution $G$ of (\ref{green1}) in two dimensional space is :
$$
\ds\partial_{\nu(y)}G(x-y,t-s)=-\frac{\nu(y)\cdot(y-x)}{4(t-s)^2}\exp(-\frac{|x-y|^2}{(t-s)}).
$$

Thus, we can derive a gain by $\nu(x)$ to get throw the formula
$\nabla(A\cdot B)=A\times(\nabla\times B)+B\times( \nabla\times
A)+(A\cdot\nabla)B+(B\cdot\nabla)A$ that :
$$
\ds\frac{\partial^2G}{\partial\nu(x)\partial\nu(y)}(x-y,t-s)=-\frac{1}{4(t-s)^2}\nu(x)\cdot\big[(y-x)\times(\nabla\times\nu(y))$$
$$+
(\nu(y)\cdot\nabla)(y-x)+\big((y-x)\cdot\nabla\big)\nu(y)+2\big(\nu(y)\cdot(y-x)\big)\frac{(x-y)}{t-s}
\big]\exp(-\frac{|x-y|^2}{(t-s)}).
$$
Therefore to achieve the proof, we insert the last formula into
relation (\ref{integ-num3}) by taking $x=\gamma(\zeta_i)$ and
$y=\gamma(\zeta_k)$. \square\\

Now, one can give an approximation to the solution $u(x,t)$ of the
problem (\ref{bvp-inv1}) by inserting the discredited formula of
the heat flux given by Theorem \ref{thm-schem1} into the
representation (\ref{poten1}) as :
\begin{equation}\label{integ-num4}
u(x_i,t_j):=hh^{\prime}\sum_{k=1}^{N}\sum_{l=1}^{N^{\prime}}G(\gamma(\zeta_i)-
\gamma(\zeta_k),t_j-t_l)\phi(\gamma(\zeta_k),t_l)|\gamma^{\prime}(\zeta_k)|
\end{equation}
\[
-hh^{\prime}\sum_{k=1}^{N}\sum_{l=1}^{N^{\prime}}\frac{\partial
G}{\partial\nu(y)}(\gamma(\zeta_i)-
\gamma(\zeta_k),t_j-t_l)g(\gamma(\zeta_k),t_l)|\gamma^{\prime}(\zeta_k)|
\]

\subsection{Measure on a smooth subset of the boundary}
Let $\Gamma\subset\subset\partial\Omega$ denote a measurable
smooth connected part of the boundary $\partial\Omega$ and
$\Gamma_c$ denotes $\partial \Omega \setminus \overline{\Gamma}$.
Introduce the trace space
$$\widetilde{\mathcal{X}} = \Bigr\{ v \in
\mathcal{X}, v \equiv 0 \mbox{ on } \Gamma_c\times(0,T) \Bigr\}.$$
Here and in the sequel we identify $g$ defined only on $\Gamma$
with its extension by $0$ to all $\partial \Omega$.\\
The aim of this section is then to identify the heat flux $\phi$
from the local measure on the Cauchy data $g\in
\widetilde{\mathcal{X}}$. To do this, we may assume that there
exists $\zeta_*\in (0,1)$ such that
$$
\Gamma:=\{\gamma(\zeta):\zeta\in[0,\zeta_*],\quad \mbox{where }
\zeta_*<<1\}.
$$
As done in last section, we introduce the subdivision of
$[0,\zeta_*]$ :
\begin{equation}\label{subdiv3}
\zeta_0=0,\zeta_i=\zeta_0+ir,i= 1,2,\cdots,M \quad\mbox{where $M$
is an integer and }r=\zeta_*/M. \end{equation} Similarly for
$[0,T]$,
\begin{equation}\label{subdiv4}
t_0=0,t_j=t_0+jr^{\prime},j= 1,2,\cdots,M^{\prime}\quad\mbox{where $M^{\prime}$ is an integer and }r^{\prime}=T/M^{\prime}.
\end{equation}
As done in Theorem~\ref{thm-schem1}, for the local measurement we
have the main result.
\begin{corollary}\label{cor-schem2}
Let $g\in \widetilde{\mathcal{X}}$ be a given function and $u\in
H^{1,1/2}(\Omega\times[0,T])$ be the solution of (\ref{bvp-inv1}).
Assume that we have the subdivision
(\ref{subdiv3})-(\ref{subdiv4}). Suppose that the heat flux
satisfy the hypothesis in Theorem \ref{thm-schem1}. Then the data
$\phi$ can be measured on $\Gamma$ as :
\begin{equation}\label{schem2}
\phi(\gamma(\zeta_i),t_j)=-rr^{\prime}\sum_{k=1}^{M}\sum_{l=1}^{M^{\prime}}\frac{\partial^2G}{\partial\nu(x)\partial\nu(y)}(\gamma(\zeta_i)-
\gamma(\zeta_k),t_j-t_l)g(\gamma(\zeta_k),t_l)|\gamma^{\prime}(\zeta_k)|
\end{equation}
\[=\frac{T\zeta_*}{4MM^{\prime}}\sum_{k=1}^{N}\sum_{l=1}^{N^{\prime}}\frac{g(\gamma(\zeta_k),t_l)}{(t_j-
t_l)^2}\frac{\gamma^{\prime}(\zeta_i)}{|\gamma^{\prime}(\zeta_i)|}\cdot\big[-\gamma^{\prime}(\zeta_k)+
\]
\[
2\big(\gamma^{\prime}(\zeta_k)\cdot(\gamma(\zeta_k)
-\gamma(\zeta_i))\big)\frac{(\gamma(\zeta_i)-\gamma(\zeta_k))}{t_j-t_l}\big]\exp(-\frac{|\gamma(\zeta_i)-\gamma(\zeta_k)|^2}{(t_j-t_l)}),
\]

for  $i= 1,2,\cdots,M;\quad j= 1,2,\cdots,M^{\prime}$ and $\gamma$
is given by Section 2.
\end{corollary}
\proof Consider that $u$ is the solution of the inverse heat
problem (\ref{bvp-inv1}), then by (\ref{integ-num1}), we can write
$$
\phi(x,t)=-\int_{0}^{t}\int_{\Gamma\cup\Gamma_c}\frac{\partial^2
G(x-y,t-s)
}{\partial\nu(x)\partial\nu(y)}g(y,s)~d\sigma(y)ds,\quad \mbox{for
} (x,t)\in\Gamma\times[0,T].
$$
The fact that $g\in \widetilde{\mathcal{X}}$ we can reduce
$$
\phi(x,t)=-\int_{0}^{t}\int_{\Gamma}\frac{\partial^2 G(x-y,t-s)
}{\partial\nu(x)\partial\nu(y)}g(y,s)~d\sigma(y)ds,\quad \mbox{for
} (x,t)\in\Gamma\times[0,T].
$$
By change of variables and by considering the subdivisions
(\ref{subdiv3})-(\ref{subdiv4}) one can deduce from the proof of
Theorem \ref{thm-schem1} that
$$
\phi(\gamma(\zeta_i),t_j)=-rr^{\prime}\sum_{k=1}^{M}\sum_{l=1}^{M^{\prime}}\frac{\partial^2G}{\partial\nu(x)\partial\nu(y)}(\gamma(\zeta_i)-
\gamma(\zeta_k),t_j-t_l)g(\gamma(\zeta_k),t_l)|\gamma^{\prime}(\zeta_k)|.
$$
Then, the proof achieves by inserting in last relation the
possible normal derivations of the function $G$. \square

\subsection{Numerical examples}
Numerical evaluations of the heat flux $\phi(x,t)$ are now
obtained by solving the integral equations (\ref{poten1-inv1})
and (\ref{integ-num1}) with the specific parameterization and discretisation of the boundary.\\

In this section, we assume that the curve is parameterized by
$\gamma(\zeta)=(\cos(2\pi\zeta),\sin(2\pi\zeta))$ and we suppose
that $T=10$, $\zeta_*=10^{-2}$ .
\subsubsection{Measure on $\partial\Omega\times[0,T]$}
We consider the following direct problem :
 \be\label{bvp-inv1-1}
\begin{array}{cc}
  \partial_tu-\Delta u=0, & (x,t)\in \Omega\times(0,T) \\
  u(x,0)=0, & x\in\Omega \\
  u(x,t)=2|x|\cos(3t), & (x,t)\in\partial\Omega\times(0,T) \\
  \frac{\partial u}{\partial\nu}(x,t)=\phi(x,t), & (x,t)\in\partial\Omega\times(0,T). \\
\end{array}
\ee We solve this problem by using Theorem \ref{thm-schem1} given
by boundary integral method which is presented in sections 3 and
4. The parameter is chosen as $N = 50$, $N^{\prime} = 100$. A
numerical result is shown in Figures.......

\subsubsection{Measure on $\Gamma\times[0,T]$}
For the case of smooth subset $\Gamma$ we consider the following
problem :
\be\label{bvp-inv1-2}
\begin{array}{cc}
  \partial_tu-\Delta u=0, & (x,t)\in \Omega\times(0,T) \\
  u(x,0)=0, & x\in\Omega \\
  u(x,t)=2|x|\cos(3t), & (x,t)\in\Gamma\times(0,T) \\
  \frac{\partial u}{\partial\nu}(x,t)=\phi(x,t), & (x,t)\in\Gamma\times(0,T). \\
\end{array}
\ee

This problem can be solved by using Corollary \ref{cor-schem2}.
The parameter is chosen as $M = 50$, $M^{\prime} = 100$. A
numerical result is shown in Figures.......

\subsection{Conclusion}
A boundary integral method for the two-dimensional inverse heat
conduction problem is considerably discussed. We presented a
numerical scheme for the inverse problem. The heat flux was
measured from the whole boundary and from a smooth subset of this
boundary. In this paper we restricted our selves to homogeneous
conduction problem, but the case with an external source may be
considered in a forthcoming work.


\end{document}